\documentclass[twocolumn]{article}
\usepackage{graphicx}
\usepackage{listings}
\usepackage{caption}

\begin{document}

\title{Linear-Time Poisson-Disk Patterns}
\author{Thouis R. Jones\thanks{Institut Curie, thouis.jones@curie.fr},
  David R. Karger\thanks{MIT CSAIL, karger@mit.edu}}
\date{February, 2011}

\maketitle
\begin{abstract}
  We present an algorithm for generating Poisson-disk patterns taking
  $O(N)$ time to generate $N$ points.  The method is based on a grid
  of regions which can contain no more than one point in the final
  pattern, and uses an explicit model of point arrival times under a
  uniform Poisson process.
\end{abstract}

\section{Introduction}

There is a long-standing interest in Poisson-disk patterns in the
graphics community, primarily for their use in sampling
\cite{Yellot:SCO:1983,Cook:SSI:1986,Mitchell:GAI:1987}.  There have
been many algorithms for generating such patterns.  Direct
implementation of ``dart-throwing''
\cite{Mitchell:GAI:1987,McCool:HPD:1992} produces true Poisson-disk
patterns, but is slow to converge.  Approximations from relaxation
\cite{Lloyd:AOA:1983} or tiling
\cite{Ostromoukhov:FHI:2003,Hiller:TBN:2001} can produce patterns
similar to Poisson-disk patterns more efficiently.  Recently, exact
methods taking log-linear time ($O(N \log N)$ where $N$ is the total
number of points) have been described
\cite{Dunbar:2006:ASD,Jones:2006:EGP}, as well as a method with
empirical $O(N)$ speed, but lacking a rigorous proof of this
performance\cite{White:PDP:2007}.

We present an algorithm with provable $O(N)$ performance.  The
algorithm maintains two data structures: a grid of regions in which
points might still be inserted, and a bucket (i.e., an unordered set)
of regions where a point {\it will} be generated (a subset of the
grid).  At each step of the algorithm, a region is taken from the
bucket, a new point is inserted in that region, and nearby regions are
updated and possibly added to the bucket.  The bucket is only empty
when no more points can be added (i.e., the pattern is {\it maximal}).
The work for each iteration is $O(1),$ for a total cost of $O(N).$

Python source code is included in the ancillary data with this paper.

\section{Background}

The Poisson Disk distribution can be defined as the limit of a uniform
two-dimensional Poisson process with a minimum-distance rejection
criterion.  Successive points are independently drawn from the uniform
distribution on $[0,1]^2$.  If a new point is at least distance $R$
from all points already accepted, it is also accepted.  Otherwise, it
is rejected.  We call this the {\it na\"ive algorithm}.  The choice of
$R$ controls the minimum distance between points (for $N$ points in
the unit square, $\pi R^2 N/4
\approx 0.548$ as $R \rightarrow 0$ \cite{Dickman:RSA:1991}).

Efficient algorithms for Poisson-disk patterns rely on generating new
points in regions where they are guaranteed (or highly probable) to be
accepted \cite{Dunbar:2006:ASD,Jones:2006:EGP,White:PDP:2007}.  In
order to guarantee equivalence of results with the na\"ive algorithm,
these methods have used $O(\log N)$ area-weighted binary search to
find where to insert a new point \cite{Dunbar:2006:ASD,Jones:2006:EGP},
or weighted spatial indexing \cite{White:PDP:2007} with theoretical
$O(\log N)$ but empirical $O(1)$ cost.

\section{Method}

\begin{figure}[t!]
\begin{center}
\includegraphics[width=0.95\columnwidth]{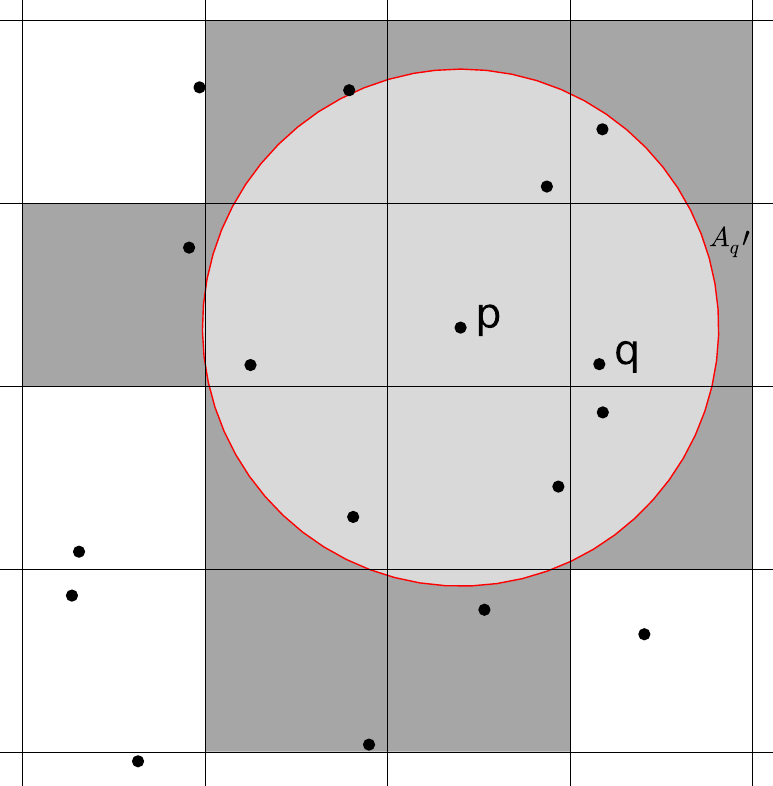}
\caption{
A point $p$ is shown with its {\it neighbor} grid squares in grey.  If $p$'s
arrival time is earlier than any of its neighbors, it will be {\it
  accepted} and added to the output.  The {\it free regions} of $p$'s
neighbors will then be updated to the dark gray areas.
This may result in points such
as $q$ being {\it invalidated}.  In $q$'s case, a replacement point
$q'$ will be generated in its new free region, $A_{q'}$.  The time of
arrival of $q'$ will be $t_q + t_+,$ where $t_+$ is drawn from an
exponential distribution parameterized by the size of the updated free region.
\label{fig1}
}
\end{center}
\end{figure}

Our algorithm can be seen as an optimization of the na\"ive algorithm
using a spatial data structure.  We store a grid with spacing $\leq
R/\sqrt{2}$ such that no more than one point can land in any grid
square in the final pattern.  We model (implicitly) a uniform 2D
Poisson process on $[0,1]^2,$ with rate $\lambda=1$, by storing at
each grid square the location and time of the earliest point landing
in that square.  These points and their arrival times will be updated
as the algorithm progresses.

Each grid square has three associated pieces of data: the free region
within that square where new points might be generated, a random point
within that free region, and a time-of-arrival for that point under a
the Poisson process.  Initially, the free region for each grid square
is the entire square, each point is chosen uniformly within its
square, and the times of arrival are drawn from $A_0 e^{-A_0 t}$,
where $A_0$ is the area of a grid square.

We also define a {\it neighbor} relationship from points to grid
squares, where the neighbors of a point are any grid squares within
$R$ of the point (see figure \ref{fig1}).

The first insight of our paper is that any point $p$ that has
time-of-arrival $t_p$ lower than any of its neighbors can be added to
the output immediately, as this indicates that $p$ arrives before any
other point that could prevent it from being accepted.  On acceptance,
the free regions of $p$'s neighbors are updated (see figure
\ref{fig1}).

It is possible that accepting $p$ will invalidate a point $q$ from
another grid square with $||p-q||_2 < R$ and $t_q > t_p$, in which
case $q$ is removed from the grid and a new point $q'$ in the updated
free region is created with a new and later $t_{q'}$ (see figure
\ref{fig1}).

The second key insight of our algorithm is that the new $t_{q'}$
should be $t_q$ plus a random variable drawn from the exponential
distribution parameterized by the area of the updated free region
$A_{q'}$, i.e., $t_{q'}=t_q + t_+$, where $t_+$ is drawn from $A_{q'}
e^{-A_{q'}t_+}$.

The logic is as follows.  The points $p$ and $q$ represent the first
arrivals in their respective grid squares, at times $t_p$ and $t_q$ (in
the na\"ive algorithm).  When $q$ is invalidated by $p$, the time
until another point arrives (in the now smaller free region) is
modeled by an exponential process with parameter $A_{q'}$.

To track which points are candidates for acceptance, we traverse the
grid and identify every point that has a time of arrival earlier than
any of its neighbors (ignoring neighbors that have already had their
points accepted).  We term these points {\it locally early}, and add
them to a bucket (an unordered set).  At each iteration, we can take
any point from the bucket, and add it to the output pattern.

Accepting $p$ may lead to new points becoming locally early, which are
then added to the bucket.  Likewise, if a point $q$ is invalidated by
$p$'s acceptance, points with $q$'s grid square in their neighbors may
become locally early, as $q$'s replacement $q'$ will have $t_{q'} >
t_q$.

\begin{figure}
\begin{center}
\includegraphics[width=0.95\columnwidth]{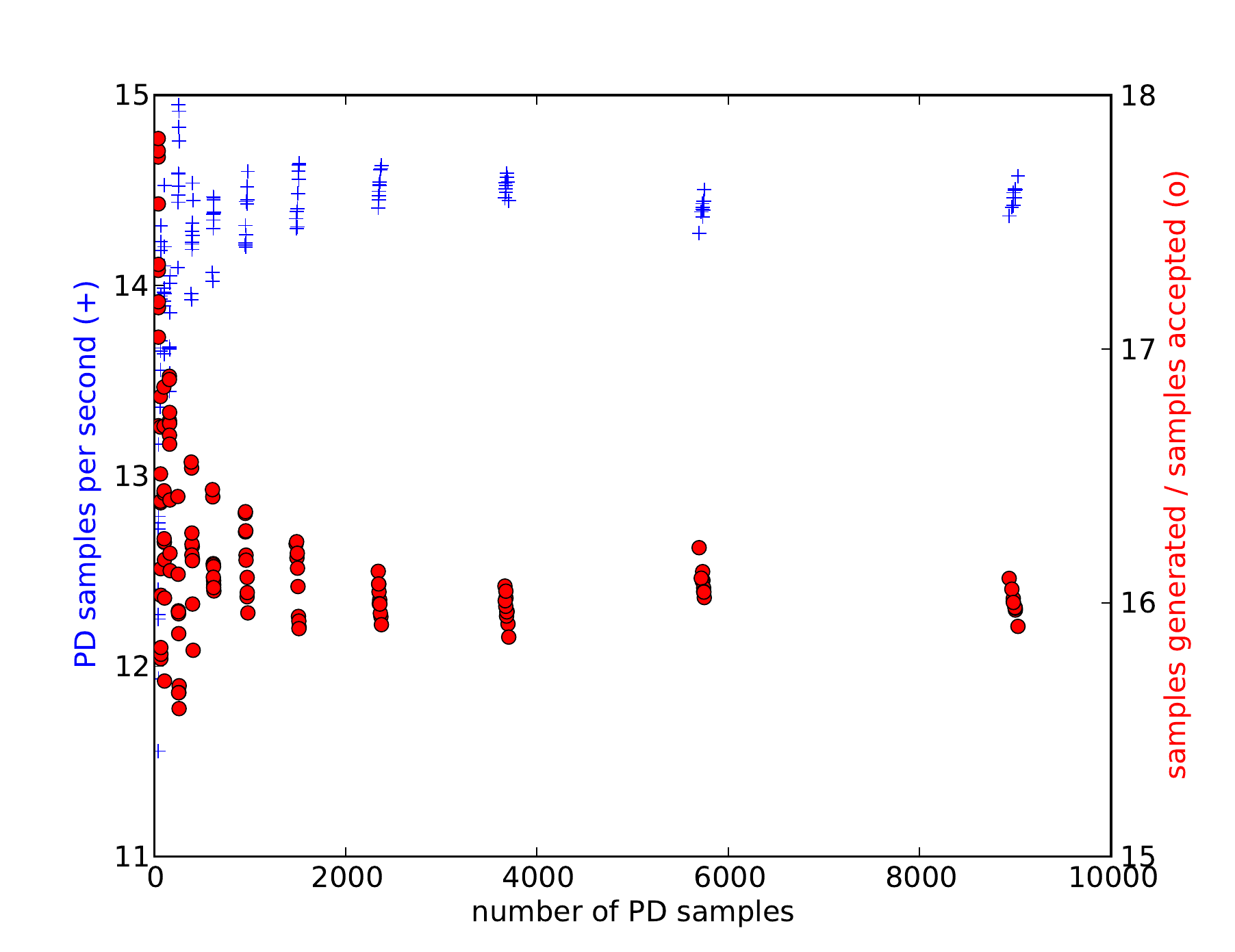}
\caption{Performance (PD samples per second) over a range of exclusion
  radii 
  (0.64 to 0.004 in a geometric progression with ratio 0.8) within the
  unit square.  The 
  number of samples generated by the Poisson arrival process versus
  the number accepted Poisson Disk samples is also shown.
\label{fig2}
}
\end{center}
\end{figure}

Each iteration is $O(1)$ provided we can update, compute the area of,
and sample uniformly from the free space of a grid square in $O(1)$
time.  Previous work has demonstrated specialized data structures
\cite{Dunbar:2006:ASD} for exactly these purposes.  In our reference
implementation, we use a constructive planar geometry library and
approximate disks with polygons for simplicity but without a loss of
generality.  We show the performance of our algorithm in terms of
samples per second, as well as number of samples generated by the
uniform Poisson arrival process versus accepted Poisson Disk samples
(see figure \ref{fig2}).  Since the size of grid squares is determined
by the radius of the PD samples, the geometric complexity of the free
space is $O(1)$.

%  In practice, rather than filling the grid in a first pass, we can
%  generate points and arrival times on demand.  We traverse the grid
%  seeking locally early points, adding them to the output immediately.
%  This leads to more points becoming locally early and added to the
%  bucket.  When the bucket is empty, the traversal continues where it
%  left off.  This leads to more coherent access and lower memory
%  requirements in the algorithm.

\section{Discussion}

We have introduced an algorithm for generating Poisson-disk patterns
in provable $O(1)$ time per generated sample.  Our main insight,
compared to recent $O(\log N)$ per point algorithms, is that rather
than choosing the location for the next point based on area-weighted
binary search, we can use an area-parameterized exponential
distribution to order points in time under a uniform Poisson arrival
process.  While previous algorithms generate each point in sequence,
with an implicit time linked to their sequential generation, we create
many points with explicit arrival times and order them (in a local
fashion) to find those that should be accepted.

\section*{Acknowledgements}
The authors wish to thank Ron Perry, Peter-Pike Sloan, and the MIT
CSAIL Computer Graphics Group for helpful comments on this paper.

\bibliographystyle{alpha}
\bibliography{paper}

\end{document}